\documentclass[useAMS,usegraphicx]{mn2e}

\usepackage{aas_macros}

\def\gs{\mathrel{\raise0.35ex\hbox{$\scriptstyle >$}\kern-0.6em
\lower0.40ex\hbox{{$\scriptstyle \sim$}}}}
\def\ls{\mathrel{\raise0.35ex\hbox{$\scriptstyle <$}\kern-0.6em
\lower0.40ex\hbox{{$\scriptstyle \sim$}}}}

\title[WISE-UKIDSS-SDSS quasars]{IR-derived covering factors for a large sample of quasars from WISE-UKIDSS-SDSS }
\author[I.G.~Roseboom et al.]
{\parbox{\textwidth}{\raggedright I.G.~Roseboom,$^{1}$\thanks{E-mail: \texttt{igr@roe.ac.uk}}, A.~Lawrence$^{1}$, M.~Elvis$^{2}$, S.~Petty$^{3}$, Yue~Shen$^{2}$, H.~Hao$^{2,4}$  }
\vspace{0.4cm}\\
\parbox{\textwidth}{\raggedright $^{1}$Institute for Astronomy, University of Edinburgh, Royal Observatory, Blackford Hill, Edinburgh EH9 3HJ, UK\\
$^2$Harvard-Smithsonian Center for Astrophysics, 60 Garden St. Cambridge MA 02138 USA\\
$^3$Physics and Astronomy Department, University of California, Los Angeles, CA 90095-1547, USA\\
$^4$SISSA, Via Bonomea 265, I-34136 Trieste, Italy\\
}}
\begin{document}

\date{\today}

\pagerange{\pageref{firstpage}--\pageref{lastpage}} \pubyear{2012}

\maketitle

\label{firstpage}

\begin{abstract}
We investigate the range of covering factors (determined from the ratio of IR to UV/optical luminosity) seen in luminous type 1 quasars using a combination of data from the WISE, UKIDSS and SDSS surveys. Accretion disk (UV/optical) and obscuring dust (IR) luminosities are measured via the use of a simple three component SED model. We use these estimates to investigate the distribution of covering factors and its relationship to both accretion luminosity and IR SED shape. The distribution of covering factors ($f_C$) is observed to be log-normal, with a bias-corrected mean of $<\log_{10}f_C>=-0.41$ and standard deviation of $\sigma=0.2$. The fraction of IR luminosity emitted in the near-IR (1--5$\,\mu$m) is found to be high ($\sim40$ per cent), and strongly dependant on covering factor. 

\end{abstract}

\begin{keywords}
quasars: general, infrared: general
\end{keywords}

\section{Introduction}

There is clear evidence that all powerful AGN are surrounded by a geometrically thick distribution of optically thick material (Rowan-Robinson 1977; Lawrence \& Elvis 1982; Antonucci \& Miller 1985; Edelson, Malkan \& Rieke 1987; Elvis et al.\ 1994; Richards et al.\ 2006). However despite nearly three decades of study the exact nature of the obscuring material has remained controversial. Recognising the difficulty in maintaining a smooth and geometrically thick rotating structure, Krolik \& Begelman (1988) suggested that the material must be ``clumpy'' or filamentary in nature, a view supported by both recent resolution VLTI observations (Jaffe et al.\ 2004; Tristram et al.\ 2007) and the lack of strong 9.7$\,\mu$m silicate emission in the mid-IR (Roche et al. 1991).

 This has motivated a large range of ``clumpy'' models, which assume a torus-shaped distribution of dust clouds surrounding the central AGN (Nenkova et al. 2002; H\"{o}nig et al.\ 2006; Nenkova et al. 2008, Schartmann et al.\ 2008; H\"{o}nig et al.\ 2010; Stalevski et al.\ 2012). These models have been found to offer reasonable agreement with the observed properties of large samples of AGN selected in a variety of ways (Alonso-Herrero et al.\ 2003; Ramos-Almeida et al.\ 2009; Mor et al. 2009; Landt et al. 2010; Alonso-Herrero et al.\ 2011; Deo et al. 2011)

However these models are almost exclusively phenomenological; they fail to prescribe a physical origin for the obscuring material, and are extremely difficult to produce/maintain in real galaxies. This has long been seen as a weakness for ``torus'' models and has driven the development of physically motivated models, including; circumnuclear star-bursts resulting from mergers (Wada \& Norman (2002); Cattaneo et al. 2005; Kawakatu \& Wada 2008), warped accretion disks (Greenhill et al.\ 2003; Lawrence \& Elvis 2010; Hopkins et al.\ 2012) and accretion disk winds (Elvis et al.\ 2002; Elitzur \& Shlosman 2006).

While the coming generation of IR and mm-wavelength interferometers (e.g. MATISSE for VLTI, ALMA) should allow direct assessment of these physical models via direct imaging of the obscuring material, much can be gleaned from statistical studies of the unresolved properties of AGN. In particular the covering factor, i.e. the fraction of sight-lines to the AGN centre obscured by dust, is a potential discriminator of physical AGN models given sufficient statistics (Lawrence 1991; Treister et al.\ 2008; Lawrence \& Elvis 2010; Elitzur 2012; Hopkins et al.\ 2012).


Until recently the limited availability of large-area mid-IR imaging has meant it was not possible to compile covering factors, traditionally probed via the relationship between re-processed dust and accretion disk emission, for large samples of AGN. The Wide-Field Infrared Survey Explorer ({\it WISE}), has begun to revolutionise the study of AGN in the IR by performing a sensitive ($\sim 100-1000$ times deeper than {\it IRAS}) all-sky survey at near-to-mid IR wavelengths. Early work has already shown the potential of {\it WISE} to return very large samples of AGN; Mor \& Trakhtenbrot (2011) presented a {\it WISE}-based study of $15,928$ quasars, a factor of $\sim50$ times the largest study possible with {\it Spitzer} (Richards et al. 2006). 

Here we make use of the {\it WISE} all-sky data release (Wright et al.\ 2010), in combination with near-IR observations from UKIDSS (Lawrence et al.\ 2007), and optical photometry/spectroscopy from SDSS to study the distribution of covering factors, and its dependance on IR and UV/optical properties, of a statistical sample of the most luminous (L$_{\rm bol}>10^{46}$\,erg\,s$^{-1}$) quasars. A number of topics can be addressed with this large uniform dataset. In this paper our aim is to address only the simplest questions: what is the typical covering factor for luminous quasars? What is the distribution of these covering factors? do any existing models predict this correctly? and, finally, what is the range in mid-IR SED shapes seen? In \S\ref{sec:data} we describe the construction of our quasar sample, while \S\ref{sec:seds} describes our SED fitting approach. \S\ref{sec:results} presents our results on the distribution of covering factors and the relationship between covering factor and IR SED shape. Finally \S\ref{sec:conc} presents our conclusions.

Throughout we assume a $\lambda$CDM cosmology with parameters $\Omega_{\rm M}=0.3$, $\Omega_{\rm \Lambda}=0.7$ and $H_0=70$\,km\,s$^{-1}$\,Mpc$^{-1}$.

\section{Data}\label{sec:data}
The parent dataset for this study is the SDSS DR7 QSO catalogue (Schneider et al. 2010) as presented by Shen et al. (2011; henceforth S11). This catalogue presents derived quantities, such as emission line fluxes, bolometric luminosities and BH masses for 105,783 Type 1 quasars brighter than $M_i=-22.0$ with reliable redshifts. To avoid concerns about host galaxy contamination we restrict the luminosity range of our study to L$_{\rm bol}>10^{46}$\,erg\,$s^{-1}$. In addition we only select quasars with {\bf $z<1.5$}, as beyond this redshift we have little information about the rest-frame mid-IR (at $z=1.5$ the {\it WISE} 22$\,\mu$m band corresponds to rest-frame 8.8$\,\mu$m). This reduces our parent sample to 26,927 quasars.

To build the {\it WISE}-UKIDSS-SDSS (WUS) quasar sample we cross-match these 26,927 quasars with first the UKIDSS Large Area Survey (LAS; Lawrence et al. 2007), then the {\it WISE} all-sky data release (Wright et al. 2011).

The UKIDSS LAS is surveying 4000 sq. deg. of the sky in four near-IR bands; $Y, J, H$ and $K$. The UKIDSS project is defined in Lawrence et al.\ (2007). UKIDSS uses the UKIRT Wide Field Camera (WFCAM; Casali et al, 2007). The photometric system is described in Hewett et al (2006), and the calibration is described in Hodgkin et al. (2009). The pipeline processing and science archive are described in Irwin et al (2009, in prep) and Hambly et al (2008). We make use of the ninth data release (DR9) which has limiting magnitudes of 20.2, 19.6, 18.8 and 18.2 mags (Vega) at $Y, J, H$ and $K$, respectively. The SDSS QSO sample is cross-matched to the LAS using a 2 arcsec search radius. Of the 26,927 quasars in our parent sample, 9230 have a counterpart in UKIDSS LAS with a detection in at least one near-IR band. 

The WISE final data release consists of imaging of the entire sky in 4 near-to-mid IR bands centred on 3.4, 4.6, 12. and 22$\,\mu$m to a depth of 0.04, 0.06, 0.5, and 3.2 mJy (3$\sigma$). We consider {\it WISE} sources which are within 3 arcsec (one FWHM for {\it WISE} at 3.4$\,\mu$m) of UKIDSS-SDSS quasars to be reliable matches. This returns 9112 matches with a detection in at least one {\it WISE} band; there remain 118 UKIDSS-SDSS quasars which do not have WISE counterparts. To ensure reliable SED fitting in the IR we only select the 5281 quasars with reliable photometry in all four {\it WISE} bands  ($>3\sigma$ and {\sc cc\_flags} equal to '0000'). For the 3831 quasars without four band {\it WISE} detections; 266 are excluded due to corrupted photometry ({\sc cc\_flags} not equal to '0000') and are not considered in the following analysis; 3565 are genuinely undetected by {\it WISE} in at least one band (typically 22$\,\mu$m) and for these upper limits ($3\sigma$) to the {\it WISE} photometry are estimated. These quasars, in addition to the 118 with no {\it WISE} counterpars, will become important when we consider the selection biases of our study.

In addition to the WUS quasars we also make use of SDSS quasars with IR photometry at 3.6, 4.5, 5.8, 8.0 and 24$\,\mu$m from {\it Spitzer} collated by Richards et al. (2006; henceforth R06). We supplement the SDSS and {\it Spitzer} photometry presented by R06 with near-IR photometry at $J$ and $K$ from the UKIDSS Deep Extragalactic Survey (DXS) and 12$\,\mu$m imaging from {\it WISE}. While the bulk of the R06 sample is detected in the {\it WISE} bands we do not use the photometry in the 3.4, 4.6 and 22$\,\mu$m channels as it overlaps with the superior {\it Spitzer} photometry at these wavelengths. Of the 259 quasars present in the original R06 catalogue, 116 have UKIDSS DXS and {\it WISE} 12$\,\mu$m counterparts within 2 arcsec. While the R06 sample is much smaller than our new WUS sample, it does have the benefit of significantly deeper ($\sim10\times$) mid-IR data from {\it Spitzer}. Again this will be important when considering the biases of the WUS sample.

\section{Estimating AGN luminosities}\label{sec:seds}
Our aim is to use simple but reliable estimates of the accretion disk luminosity and the fraction of this re-processed by dust and emitted in the IR. We do this by fitting SED models as described below. We do not require that the precise parameters of these model fits to be meaningful, as our SEDs have only a limited number of photometric bands. Rather, they are simply intended as an objective way to characterise the data.

S11 present bolometric accretion disk luminosities for all of our WUS quasar sample, while R06 similarly present accretion disk luminosities (henceforth refered to as $L_{\rm bol}$) for that sample, and we make use of these here.  

\begin{figure}
\includegraphics[scale=0.35]{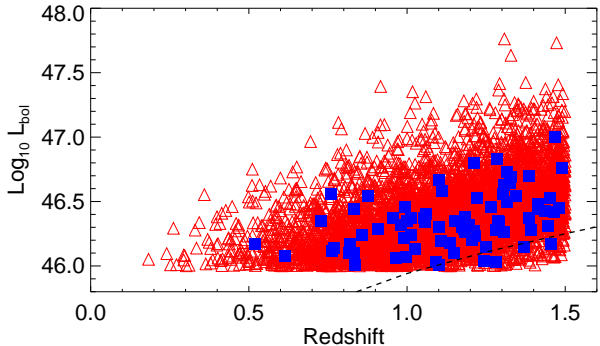}
\caption{Luminosity ($L_{\rm bol}$) vs. redshift for the WUS (red triangles) and R06 (blue filled squares) samples. The dashed black line shows the effective luminosity limit for quasars with a E94 SED imposed by the $i<19.1$ magnitude limit of the parent SDSS quasar sample. It is clear that the depth of the IR imaging used here is sufficient to probe the full range in $z-L_{\rm bol}$ sampled by the SDSS QSO dataset.  }
\label{fig:lz}
\end{figure}

Fig.~\ref{fig:lz} shows $L_{\rm bol}$ as a function of redshift for both the WUS and R06 samples. The median luminosity and redshift of the WUS and R06 samples are comparable, with values of $L_{\rm bol}^{\rm WUS}=46.4$ and $z^{\rm WUS}=1.1$ for WUS and medians of $L_{\rm bol}^{\rm R06}=46.2$ and $z^{\rm R06}=1.0$ for the R06 sample. It can be seen that {\it WISE} is able to detect SDSS quasars above $10^{46}\,L_{\rm bol}$ out to at least $z\sim1$ and is not significantly biased to high $L_{\rm bol}$ when compared to the deeper R06 sample.

\begin{figure}
\includegraphics[scale=0.35]{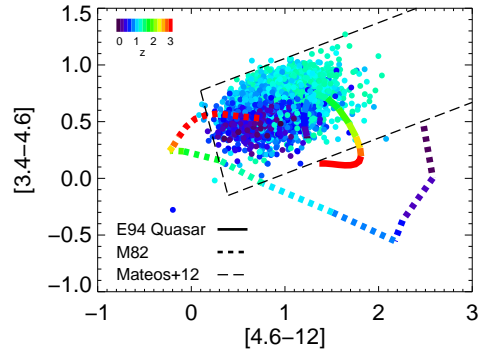}
\caption{{\it WISE} colours [3.4-4.6] vs. [4.6-12] for WUS quasars (AB mags). Overplotted are tracks taken by a quasar mid-IR SED (E94; solid line) and the local starburst galaxy M82 (dashed line). The colours of the WUS quasars show good agreement with the predicted E94 {\it WISE} colours, and are clearly separated from the predicted colours of M82. }
\label{fig:wuscc}
\end{figure}

For the IR luminosities we integrate the predicted flux density from the best-fit three component SED model. The three components of our SED model are; the accretion disk, a ``hot'' dust component, and an obscuring torus. No IR component from star formation is considered, meaning contamination in the mid-IR from star formation is effectively ignored. We justify this omission in two ways. Firstly, the observed {\it WISE} colours of our quasar sample are uniformly consistent with AGN dominated SEDs. In Fig.~\ref{fig:wuscc} we show the {\it WISE} colours [4.6-12] vs. [3.4-4.6] (AB mags) for our WUS sample, compared to the predicted location of quasar (E94) and the starburst (M82) SEDs, and the AGN colour selection ``wedge'' of Mateos et al.\ (2012). The WUS quasars show good agreement with the E94 quasar track, with the vast majority located within the AGN selection wedge. The {\it WISE} colours of WUS quasars are also clearly separated from the M82 track, very few of the WUS quasars lie close to the predicted colour evolution of M82 and in those cases the redshifts disagree significantly (i.e. $z>3$ starbursts show similar colours to $z\sim0.1$ quasars). The results of Fig.~\ref{fig:wuscc} strongly suggest that the mid-IR SEDs of WUS quasars are relatively unaffected by star formation. This is not surprising if we consider the luminosity of the WUS sample. The minimum luminosity of our WUS QSO sample is $L_{\rm bol}=10^{46}$\,erg\,s$^{-1}$, for IR emission from star formation to be comparable to this would require a star formation rate (SFR) of $\sim 450$\,M$_{\odot}\,$yr$^{-1}$ (assuming the translation between SFR and IR luminosity given by Kennicutt 1998). While this is feasible, indeed submm/mm observations of QSOs have revealed some host SFRs of 100--1000s\,M$_{\odot}$\,yr$^{-1}$ (Isaak et al. 2002; Omont et al. 2003; Beelen et al.\ 2006; Hatziminaoglou et al.\ 2010; Dai et al.\ 2012), the typical SED for a star forming galaxy has an effective grey body temperature of $\sim40$\,K (compared to $\sim1500$\,K for AGN tori) and thus only a small fraction of the total IR luminosity is emitted at the mid-IR wavelengths probed by {\it WISE}. To quantify this, at the maximum redshift of our WUS sample ($z<1.5$) the longest wavelength probed is $\sim9\,\mu$m. For the M82 SED used in Fig.~\ref{fig:wuscc} only 15 per cent of the IR luminosity is contained in the wavelength range 1--9$\,\mu$m (compared to $\gs60$ per cent for the N08 AGN torus models considered here). 

At the minimum quasar luminosity we consider ($10^{46}L_{\rm bol}$), for star formation to contribute 0.2\,$L_{\rm bol}$ to the observable mid-IR luminosity in the {\it WISE} wavebands would require a SFR of 600\,M$_{\odot}$\,yr$^{-1}$, while for the mean luminosity of the WUS sample ($\langle L_{\rm bol}\rangle=46.4$), this would increase to $\sim2000$\,M$_{\odot}$\,yr$^{-1}$. The question remains, are these SFRs common for quasar hosts at $z<1.5$? Recently {\it Spitzer} and {\it Herschel} have provided far-IR observations for large samples of quasars for the first time (Shi et al.\ 2009; Hatziminaoglou et al.\ 2010; Bonfield et al.\ 2011; Dai et al.\ 2012). There is some evidence of a relationship between $L_{\rm bol}$ and SFR (as probed by $L_{\rm FIR}$), with the form of $L_{\rm FIR}\propto L_{\rm bol}^{0.3\pm0.1}$ (Hatziminaoglou et al.\ 2010; Bonfield et al.\ 2011). Taking the results of Hatziminaoglou et al.\ (2010; Fig.~3) we can see that for submm detected type 1 AGN (which form only 15 per cent of the total population) $\log_{10}L_{\rm FIR}\sim 45-46$\,erg\,s$^{-1}$ at $L_{\rm bol}=10^{46}$\,erg\,s$^{-1}$, equivalent to a SFR of between 45-450\,M$_{\odot}$\,yr$^{-1}$. Given that these sources must represent the most star forming quasars (as $\sim85$ per cent of quasars remain undetected at {\it Herschel} wavelengths) the large SFRs needed to significantly affect the mid-IR SEDs considered here must be extremely rare in quasar hosts.

We now describe each of the model components in detail. For the accretion disk component we assume a single template; the Elvis et al.\ (1994) radio-quiet quasar template from $0.05<\lambda<0.7\,\mu$m extrapolated smoothly with a power-law of slope $\lambda f_{\lambda}\propto\lambda^{-1}$ at $\lambda>0.7\,\mu$m (Elvis et al. 1994; Hao et al. 2010). Emission from the H$\alpha$ line at 6563\AA\/ is also incorporated with a single Gaussian with $\sigma=30$\AA\/ and a EW$=20$. We also consider extinction of the accretion disk by dust, assuming SMC-like extinction curve (Prevot et al. 1984). In all cases the multi-component fit is made using the {\sc IDL} routine {\sc MPFIT} (Markwardt 2008). The IR-luminosities are calculated by integrating the best-fit SED, taking the hot-dust and ``torus'' components only in the range $1<\lambda<1000\,\mu$m.

For the hot dust we assume a simple blackbody described by two parameters;  the amplitude and temperature. This component is added as previous work has found it difficult to match the observed IR properties of quasars with a single clumpy torus (Mor et al. 2009). To avoid degeneracies with the accretion disk and dust torus emission we restrict the temperature of the BLR hot dust to lie in the range $100<T<2000\,$K. This upper limit is also consistent with the sublimation temperature for graphite grains ($\sim1900\,$K; Barvainis 1987)

Finally, the dust torus emission is characterised by one of three template SEDs selected from the clumpy torus models of Nenkova et al.\ (2008; henceforth N08)\footnote{\url{http://www.pa.uky.edu/clumpy/}}. While the full range of the N08 models could have been used (N08 present over 100,000 different realisations of their model parameters) it is unlikely that the {\it WISE} 4-band IR photometry can distinguish between these models. Moreover, it is not computationally feasible to fit the full sample of over 5000 WUS quasars with such a detailed model. Given that the intention here is to use the models to interpolate between the observed data points, not to interpret physical properties of individual quasars, a small subset (i.e. three) of torus models should be sufficient.

To confirm this, and to determine which template(s) to use, we selected a representative sample of 6000 N08 templates and fit them to the smaller R06 quasar sample. The templates were selected on a representative grid in terms of the N08 model parameters; viewing angle ($i$), optical depth ($\tau_V$), number of clouds ($N_0$), opening angle ($\sigma$), power-law index of the radial distribution ($q$) and ratio of the external to internal radii ($Y$). The details of the N08 model grid are given in Table \ref{tab:n08grid}. The R06 sample is used as a testbed, rather than the WUS sample, as it has very similar characteristics (similar wavelength sampling, similar redshift and luminosity range) to the WUS sample, but is almost ten times deeper in the mid-IR, and so is not biased to mid-IR bright SED shapes. For each template on the grid we fitted the R06 sample using the full SED model described above.

\begin{table}
\caption{Details of N08 model grid used to fit R06 quasars}
\label{tab:n08grid}
\begin{tabular}{lll}
Parameter & N & Values\\
\hline\hline
$i$ & 5 & 0, 20, 40, 60, 80\\
$\sigma$ & 5 & 15, 30, 45, 60, 75\\
$Y$ & 4 & 10, 30, 60, 100\\
$N_0$ & 4 & 3, 6, 9, 12\\
$q$ & 3 & 0, 1, 2\\
$\tau_V$ & 5 & 20, 40, 60, 80, 100\\
\hline
\end{tabular}
\end{table}


On the basis of these SED fits, we use a K-means clustering algorithm (Lloyd 1982) to group the R06 quasars by which N08 model provides the best-fit. The algorithm consists of iterating over two steps. In the first step each quasar is allocated to the group which offers the lowest $\chi^2$. In the second step the best model for each group is determined by finding the N08 template which offers the lowest mean $\chi^2$. These steps are repeated until convergence (group assignments remain the same for two consecutive iterations). The initial templates for each group were chosen to be roughly representative of the spread in best-fit N08 parameter values seen in the R06 samples. We run the algorithm assuming 2, 3, 4 and 5 groups. The best single N08 torus model (i.e. one group) offers a mean $\chi^2=22.4$. The mean $\chi^2$ across the R06 sample reduces from $\langle\chi^2\rangle=18.7$ in the two model case to $\langle\chi^2\rangle=17.9$, $\langle\chi^2\rangle=17.6$ and $\langle\chi^2\rangle=17.4$ for the 3, 4 and 5 model cases respectively. Given these results we justify the use of the three N08 templates to represent the WUS quasars; using more N08 models does not decrease the mean $\chi^2$ considerably.

\begin{table}
\caption{Details of three N08 templates used to fit the quasar samples}
\label{tab:N08mods}
\begin{tabular}{lllllll}
\hline\hline
 & $i$ & $\sigma$ & $Y$ & $N_0$ & $q$ & $\tau_V$ \\
\hline
1 & 0 & 30 & 100 & 3 & 1 & 20\\
2 & 0 & 60 & 30 & 3 & 1 & 20\\
3 & 20 & 60 & 100 & 9 & 0 & 20\\
\hline
\end{tabular}
\end{table}

The details of the three N08 models chosen by this process are given in Table \ref{tab:N08mods}. The parameter values we derive for the representative model set are noteworthy in two ways. Firstly, in all three templates our fitting calls for large values of $Y$, the ratio of outer to inner radius, and $n$ the number of clouds, while requiring very low values of $\tau_V$, the optical depth of the clouds. These values are at odds with other attempts to use the N08 models to describe the {\it Spitzer} IRS spectra of PG quasars (Mor et al. 2009), which typically call for lower values of $Y$ and $\sigma$ and much larger values of $\tau_V$. Secondly, while the first model would be seen as a Type 1 AGN the majority of the time, both the second and third model would rarely be seen as a Type 1 AGN; the accretion disk should be obscured $\sim80$ and $\gs 90$ per cent of the time given the parameter values used for the two models, respectively (Elitzur 2012). For these reasons it is unlikely that the parameter values from our fitting are physically meaningful, either because of our unsophisticated fitting methodology or the limited number of mid-IR wavebands observed. However, our SED fits provide a meaningful estimate of the total IR luminosity and the gross shape of the IR SED by allowing us to interpolate between the mid-IR photometric points. Indeed, comparing the best fit SEDs from the N08 model grid to those derived just these three models the median $\chi^2$ is found to increase (median $\chi^2=11.1$ for the grid vs. $\chi^2=12.8$ for the three model case), but the derived quantities of interest (e.g. $L_{\rm IR}$ and $L_{1-5\mu m}$) do not change significantly. To quantify this, for the IR luminosity we find the mean ($\mu$) and standard deviation ($\sigma$) of $\left(L_{\rm IR}^{\rm gr}-L_{\rm IR}^{\rm 3m})\right / L_{\rm IR}^{\rm gr}$ to be $\mu=0.006$ and $\sigma=0.16$, where $L_{\rm IR}^{\rm gr}$ is the IR luminosity calculated from the best-fit to the full N08 grid, and $L_{\rm IR}^{\rm 3m}$ as the best-fit from the three templates. Similarly for $L_{1-5\mu m}$ the mean ($\mu$) and standard deviation ($\sigma$) of $\left(L_{\rm 1-5\mu m}^{\rm gr}-L_{\rm 1-5\mu m}^{\rm 3m})\right / L_{\rm 1-5\mu m}^{\rm gr}$ is found to be $\mu=0.007$ and $\sigma=0.02$. It should be noted that here (and throughout) $L_{\rm IR}$ and $L_{1-5\mu m}$ do not include any contribution from the accretion disk component. Assuming that the grid of N08 models completely describe the range of possible ``torus'' shapes, these values represent the systematic and random error introduced by our adoption of a limited template set. 

\begin{figure}
\begin{centering}
\includegraphics[scale=0.4]{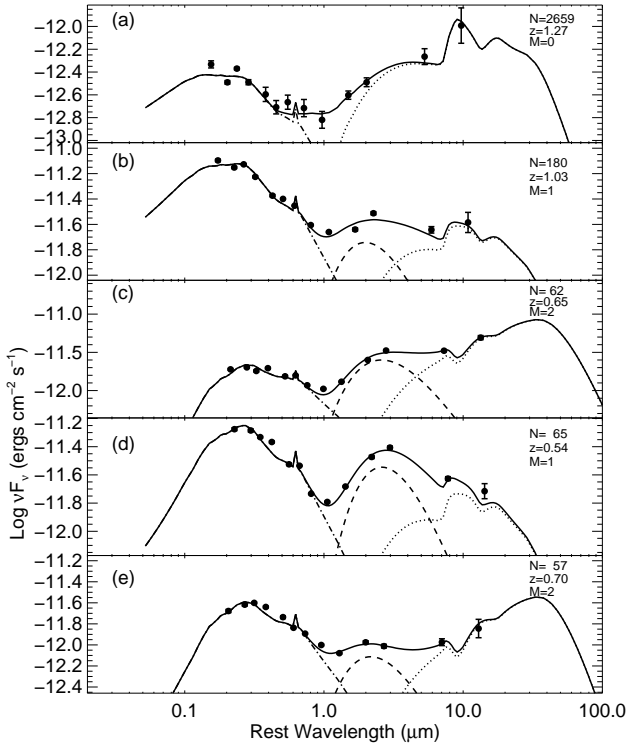}
\caption{Examples of SED fits to WISE-UKIDSS-SDSS (WUS) quasars. In each panel the solid line is our best fit SED, while the independent contributions from the accretion disk (dot-dashed line), broad line region hot dust (dashed line) and dust torus (dotted line) are also shown. The photometric data, de-redshifted, is shown as filled dots. Errors are 1$\sigma$. The panels show randomly chosen examples of (a) all with good fits ($\chi^2<20$), (b) IR-faint, (c) IR-bright , (d) near-IR excess WUS quasars and (e) near-IR faint WUS quasars. The number of objects falling into each category is given in the top right of each panel. It can be seen that in all cases our model prescription does an adequate job of describing the observed SEDs of WUS quasars.}
\label{fig:exfits}
\end{centering}

\end{figure}

For each WUS quasar we find the best fit combination of accrection disk, hot dust and torus components. We find the median $\chi^2$ of the WUS SED fits to be $\chi^2=19$, slightly higher than the R06 sample despite the number of degrees of freedom being equal (Seven; thirteen bands less six free parameters). Of the 5281 WUS QSOs with reliable {\it WISE} photometry; 3600 are best fit by model (1) in Table \ref{tab:N08mods}, 1027 by model (2) and 654 by model (3). Examples of five SED fits which roughly span the range of SEDs seen in the WUS sample are shown in Fig.~\ref{fig:exfits}.

It can be seen from Fig.~\ref{fig:exfits} that, despite its simplicity, our SED modelling methodology has the ability to reproduce the observed SEDs of a wide variety of WUS quasars with reasonable precision. Fig.~\ref{fig:exfits} shows randomly chosen examples of (a) all WUS quasars with good fits ($\chi^2<20$), (b) IR-faint (covering factor less than one-fifth), (c) IR-bright (covering factor greater than one), (d) near-IR excess ($L_{1-5\mu m}/L_{\rm IR}>0.55$) and (e) near-IR faint WUS quasars ($L_{1-5\mu m}/L_{\rm IR}<0.2$). As well as illustrating the effectiveness of our SED fitting, these examples show that there is a significant range in the relative strength of the near-IR, mid-IR and accretion disk emission in the quasar population probed by WUS.

\section{Results}\label{sec:results}
\begin{figure}
\begin{centering}
\includegraphics[scale=0.45]{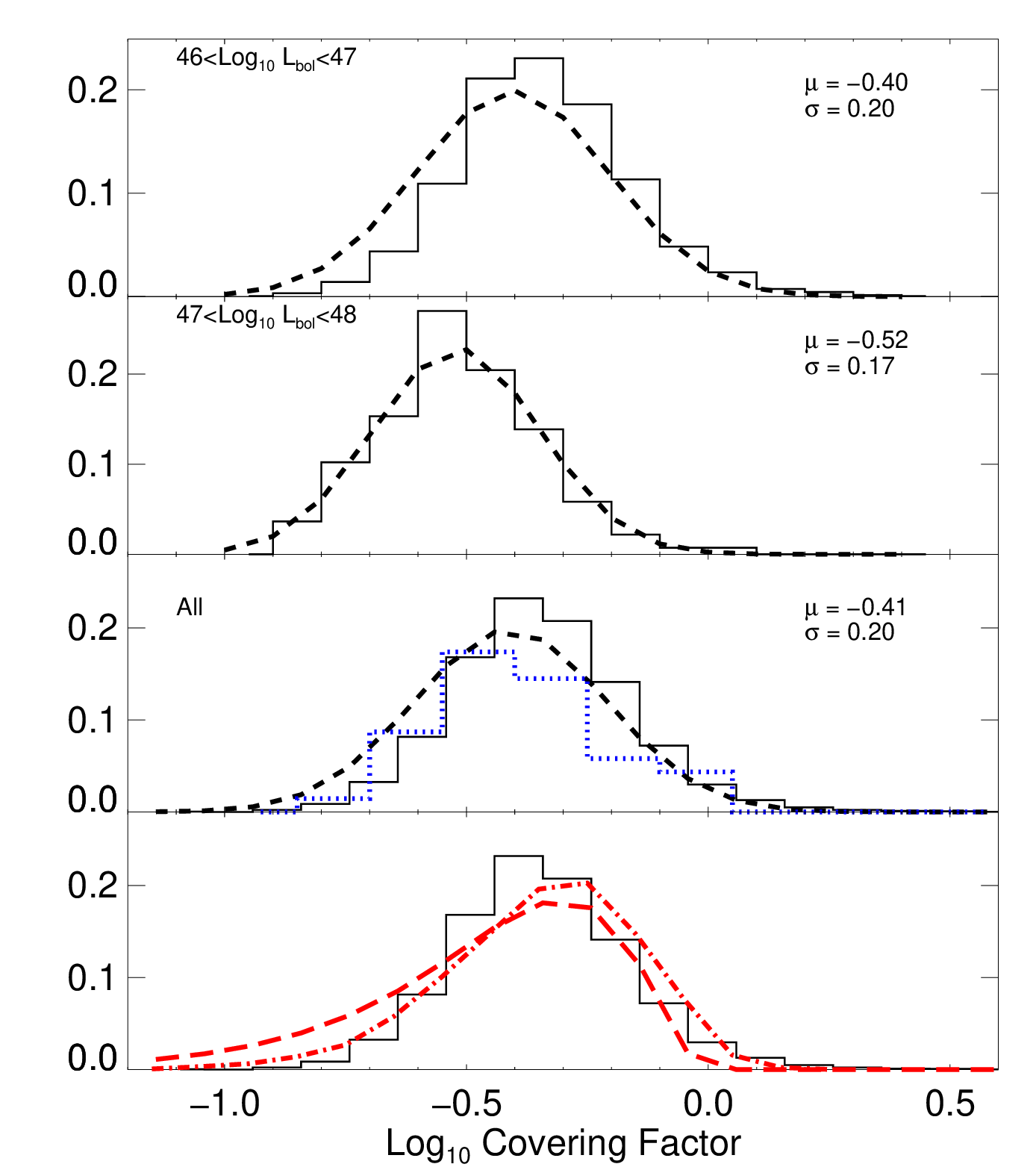}
\caption{Distribution of covering factor ($f_C$; defined as L$_{\rm IR}/$L$_{\rm bol}$) for WUS quasars. The solid line shows the binned histogram of $f_C$'s for WUS quasars with good ($>3\sigma$) {\it WISE} detections in all four bands. The dashed line shows the best-fit Log-Normal distribution to the full WUS sample, taking into account upper limits for those quasars without 4-band {\it WISE} detections. The top two panels show the distributions in bins of $\log_{10} L_{\rm bol}$, while the bottom two panels shows the distribution for all WUS quasars. Also shown in the third panel is the distribution for the R06 sample (blue dotted line). In the bottom panel the distribution of $f_C$ for Type 1 quasars is shown as a red long-dashed (dot-dashed) line for the warped accretion disk model of LE10 uncorrected (corrected) for the selection limits of {\it WISE}.}
\label{fig:cfs}
\end{centering}

\end{figure}

The ratio of the IR to the bolometric luminosity is commonly interpreted as a ``covering factor'', i.e. the fraction of sight-lines to the accretion disk which are obscured (Maolino et al. 2007; Hatziminaoglou et al.\ 2008; Rowan-Robinson et al. 2009). Fig.~\ref{fig:cfs} shows the distribution of covering factors ($L_{\rm IR}/L_{bol}$) for the WUS quasars. The mean (and standard deviation) of the observed covering factors is $\langle \log_{10} f_C^{\rm obs} \rangle=-0.34$ and $\sigma_{\log_{10}f_C}^{\rm obs}=0.19$. To account for the significant fraction ($\sim40$ per cent) of quasars for which we only have limits to $L_{IR}$, we assume a log-normal distribution for the data, and use the expectation-maximisation (EM) algorithm for maximum likelihood estimation of censored data presented by Wolynetz (1979). This results in a corrected mean and standard deviation of $\langle\log_{10}{f_C}\rangle=-0.41$ (i.e. $\langle f_C\rangle=0.3$) and $\sigma_{\log_{10}f_C}=0.2$. We can test our assumption of a log-normal distribution for the covering factor in two ways. First we can compare our bias corrected distribution to that seen in the smaller, but deeper, R06 quasar sample. The mean (and standard deviation) of the covering factors in the R06 sample is $\langle \log_{10}f_C^{\rm R06} \rangle=-0.41$ and $\sigma_{\log_{10}f_C}^{\rm R06}=0.19$, almost identical to the corrected WUS values. As second test we can use a Monte-Carlo simulation to predict what the observed distribution should be assuming the corrected one, and using the known limits of the {\it WISE} photometry.

In each realisation samples are created with covering factors, redshifts and bolometric luminosities drawn from measured distributions. The detectability of each source is tested by assuming the mean mid-IR SED and predicting the observed {\it WISE} fluxes, assuming Gaussian errors consistent with the catalogued values. Sources are considered detected if they have {\it WISE} flux densities above the $3\sigma$ detection limits quoted in \S\ref{sec:data}, subject to incompleteness levels consistent with the values presented in the {\it WISE} explanatory material\footnote{\url{http://wise2.ipac.caltech.edu/docs/release/allsky/expsup/}}. This process is repeated 100 times to produce statistically robust results. In our Monte-Carlo realisations we detect on average 61.9 per cent of sources, comparable to the actual detection rate of 59.6 per cent (5281/8846). The mean and standard deviation of the recovered covering factors in the simulation is $\langle\log_{10} f_C^{\rm sim} \rangle=-0.33$ and $\sigma_{\log_{10}f_C}^{\rm sim}=0.19$, in good agreement with the values for the observed distribution in WUS quasars. 

A potential variation is seen as a function of luminosity, with the mean  $\log_{10} f_C$ decreasing by 0.11 across the two luminosity bins. Although this trend cannot be confirmed with the limited luminosity range of our sample, it is consistent with that seen in the analogous studies of {\it WISE} quasars by Mor \& Trakhtenbrot (2011) and Calderone, Sbarrato \& Ghisellini (2012). 

\begin{figure*}
\includegraphics[scale=0.4]{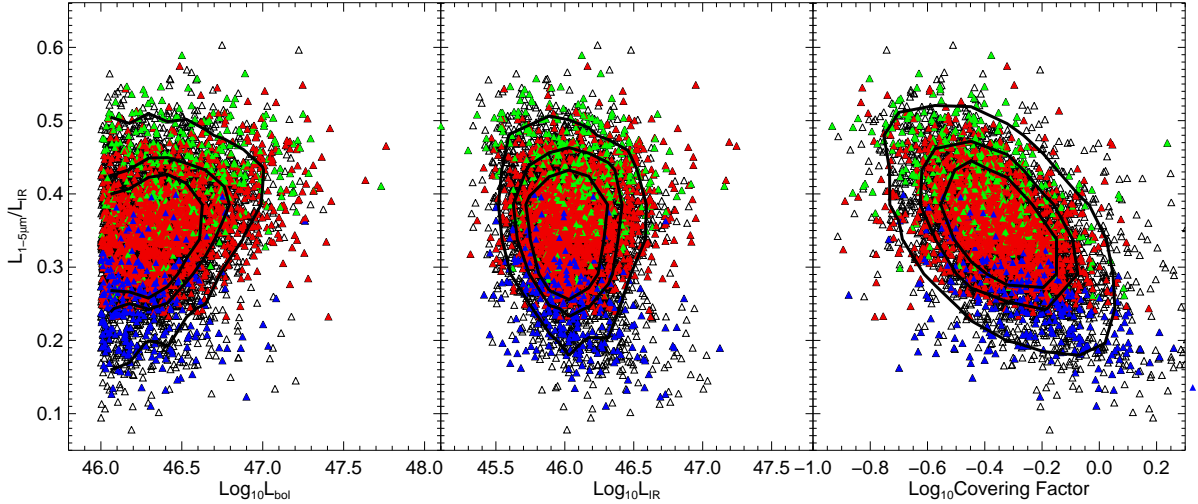}
\caption{Ratio of near-IR (1--5$\,\mu$m) to the total IR luminosity ($L_{1-5\,\mu m}/L_{\rm IR}$) as a function of; $L_{\rm bol}$ (left), $L_{\rm IR}$ (middle) and covering factor (right). Open triangles are all WUS quasars with reliable photometry, while filled triangles represent those with good SED fits ($\chi^2<20$). Filled symbols are colour-coded by the N08 model fitted (see Table \ref{tab:N08mods}); with red representing model 1, green is model 2 and blue is model 3. Black contours enclose 90, 70 amd 50 per cent of the WUS population. A strong correlation between $L_{1-5\,\mu m}/L_{\rm IR}$ and $f_C$ can be seen.}
\label{fig:lh}
\end{figure*}

As our ability to constrain the shape of the IR SED with typically only 4--6 photometric points is limited, we test variations in the IR SED shape by comparing the near-IR (1--5$\,\mu$m) to the total IR luminosity ($L_{1-5\,\mu m}/L_{\rm IR}$). Figure \ref{fig:lh} shows $L_{1-5\,\mu m}/L_{\rm IR}$ as a function of bolometric luminosity, IR luminosity and covering factor. While $L_{1-5\,\mu m}/L_{\rm IR}$ appears relatively insensitive to $L_{\rm bol}$ and $L_{\rm IR}$, a strong correlation appears between $L_{1-5\,\mu m}/L_{\rm IR}$ and $f_C$. To test the significance of this correlation we calculate the Spearman rank correlation coefficient ($\rho$), finding $\rho=-0.5$. Given the sample size this correlation has a $<<1$ per cent probability of occurring by chance.

\section{Discussion}\label{sec:disc}
\subsection{The typical covering factor for high luminosity quasars}
The simplest result is that the mean observed value of $\log_{10}f_C=$-0.41, i.e. the typical quasar is observed to have $f_C\sim$0.39. This is consistent with early SED studies such as those of Sanders et al.\ (1989) and Elvis et al.\ (1994), but seems to be inconsistent with the SDSS/Spitzer study of Treister et al.\ (2008), who found $L_{24}/L_{\rm bol}\sim$ 0.04 at the highest luminosities.

It is important to note that, under the assumption that the Type 1/2 divide is purely an orientation effect, our observed distribution of covering factors must be biased, such that $p_1(f_C)=(1-f_C)p(f_C)$, where $p_1(f_C)$ is the probability of seeing a Type 1 AGN (quasar) with covering factor $f_C$ and similarly $p(f_C)$ is the probability of seeing any AGN with covering factor $f_C$ (Lawrence 1991, Lawrence \& Elvis 2010, Elitzur 2012). Interestingly, applying this correction gives us a median covering factor for all AGN of 0.64, in good agreement with the ratio of Type 2 to Type 1 AGN seen in the IR/radio selected samples (LE2010; based on De Grijp et al. 1992; Rush et al. 1993; Lacy et al. 2007). However it is in serious disagreement with claims based on X-ray samples that at the highest luminosities the obscured fraction is only 5--10 per cent  (Hasinger et al. 2008; Tueller et al. 2008), although this extra X-ray obscuration may originate from dust-free gas clouds very close to the accretion disk (e.g. Risaliti et al. 2007; Elvis et al. 2012). 
\subsection{The distribution of covering factors}
It is striking that there is a clear distribution of covering factors, but that the distribution is fairly narrow. The dispersion in $\log_{10} f_C$ of 0.2 means that two-thirds of type 1 quasars have $f_C$ in the range 0.25--0.61. In principle the shape of the covering factor distribution is a test of models for AGN obscuration. In practice such models have not so far made such predictions. The exception is the simple tilted disk model of Lawrence \& Elvis (2010; henceforth LE10), which predicts $p(c)=\frac{1}{2}\sin\pi f_C$ (We ignore the other model proposed in LE10, the twisted disk, as it is known to give an unrealistic distribution of covering factors).

 We show this prediction for the tilted disk model in Fig.~\ref{fig:cfs}, taking into account the bias for Type 1 AGN, i.e. $p_1(c)=\frac{1}{2}(1-f_C)\sin\pi f_C$ . This very simple model, with no adjustable parameters, comes close to the observed distribution, but with predicted values slightly higher than the observed ones. This suggests that a detailed physical warped disk model may be able fit the observed distribution. It would be very interesting to see predictions from disk wind and star-burst disk models.

To determine the effect of the {\it WISE} selection limits on the LE10 tilted disk model we performed a Monte-Carlo simulation by assigning covering factors randomly selected from the LE10 tilted disk distribution to our real WUS quasars. This process was repeated 100 times and a prediction of the observed distribution made, excluding sources which would now be below the {\it WISE} detection limits and including a random error of 15 per cent on our measurements of the covering factor; this is shown in Fig.~\ref{fig:cfs}. Interestingly, our simulation returns roughly the right completeness, we would expect to detect 5185 quasars on average (similar to our actual number of 5281), and a log-normal distribution, with the peak of the distribution offset by $\log_{10}{f_C}\sim0.1$ dex from the observed distribution. In order to bring our observed distribution in line with these predictions our estimates of the covering factor must be systematically biased low by $\sim25$ per cent, or an additional absorber unassociated with the quasar must be present (e.g. galaxy scale dust; Elvis 2012).

\subsection{The relationship between covering factors and near-IR emission}
The results of Fig.~\ref{fig:lh} present us with two puzzles; is the fraction of IR luminosity emitted in the near-IR consistent with AGN models? and why is it a strong function of covering factor?

Taking these questions in turn, the typical ratio of near-IR (1--5$\,\mu$m) to total IR luminosity seen here is $\sim40$ per cent. This is close to the maximum $L_{1-5\,\mu m}/L_{\rm IR}$ achievable within the N08 models, and is only possible with a limited range of model parameter values (e.g. $Y<30$, $N\le4$, $q\ge1$, $\tau\le20$). This discrepancy is not unique to our work, several previous studies attempting to explain the IR SEDs of AGN have found the need for extra ``hot'' dust components (Polleta et al.\ 2008; Mor, Netzer, \& Elitzur 2009; Deo et al. 2011; Vignali et al.\ 2011).

 A potential reason for this excess may be the dust composition. Thermal emission peaking at $\sim1\,\mu$m require temperatures which are greater than the sublimation temperature for the silicate-type grains (T$\sim1500$\,K) that make up the bulk of the dust assumed by the models (N08; Draine \& Lee 1984). However some dust must survive very close to the accretion disk in order to explain the correlation between the optical and near-IR variability seen in nearby AGN (Minezaki et al.\ 2004; Suganuma et al.\ 2006; Kishimoto et al.\ 2007). While the single blackbody employed here is unlikely to be the correct model for the hot dust, the Wien side of the blackbody fit is well constrained by the UKIDSS and {\it WISE} near-IR data and hence we may interpret our fitted blackbody temperatures as a ``maximum'' temperature for the dust closest to the accretion disk. Encouragingly, the mean ``maximum'' temperature for WUS quasars is 1481,K, below the sublimation temperature for silicate grains, and in good agreement with the value (1400\,K) found via a similar analysis of PG quasars by Mor, Netzer \& Elitzur (2009). Only 37 per cent of WUS quasars are found to need a hot dust in excess of 1500\,K, suggesting that variations to the dust composition are not needed to explain the IR SEDs of most luminous quasars. 

The assumptions made by the ``torus'' model have a large effect on the near-IR luminosity. Models which assume a smooth, rather than clumpy, distribution of dust (e.g. Fritz et al.\ 2006) naturally result in a large fraction of the IR luminosity emitted in the near-IR (see discussion in Vignali et al.\ 2011). As pointed out by Vignali et al.\ (2011), the failure of clumpy models to reproduce the large near-IR luminosities seen here without additional hot components points to serious problems with the underlying paradigm of the ``clumpy torus''. However smooth models are not without their own failings; both in terms of their stability (Krolik \& Begelman 1988) and their ability to reproduce the range of observed 10$\,\mu$m silicate emission/absorption (Nikutta, Elitzur \& Lacy 2009).

 The recent model of Stalevski et al.\ (2012) tries to include aspects of both clumpy and smooth models by using a ``two-phase'' model; essentially a clumpy model embedded in lower-density, smoothly distributed, dust. Computing the near-IR to total IR luminosity for their template SEDs\footnote{\url{https://sites.google.com/site/skirtorus/download/description}} we find that near-IR luminosity ratios similar to that seen here (e.g. L$_{1-5\,\mu m}$/L$_{\rm IR}>0.4$) are easily achievable for these models. 

Finally, we consider the implications of the correlation between the ratio of near-IR to total IR luminosity and covering factor. A simple solution is that the near-IR (i.e. ``hot'') and mid-IR (i.e. ``warm'') components of the SED come from physically distinct components. If the covering factor of the ``hot'' component was relatively constant across all quasars, while the ``warm'' component varied, it would produce a relation similar to that seen in Fig.~\ref{fig:lh} (as the ``hot'' and ``warm'' covering factors must add up to $f_C$). However, this would require an obscurer for the``hot'' component physically distinct from the ``torus'' which is needed to explain the mid-IR emission. Given the result shown in Fig.~\ref{fig:lh}, for this scenario to work the covering factor of the ``hot'' dust component must be $\sim5-15$ per cent. Interestingly, this is close to the value assumed for the broad line region clouds ($\sim10$ per cent; Wyithe \& Loeb 2002).

Another explanation for the observed $f_C$ -- L$_{1-5\,\mu m}$/L$_{\rm IR}$ relation, which does not require two distinct obscurers, is that it is a geometric effect. As the covering factor decreases, the maximum inclination at which a type 1 quasar would be seen increases. An increase in the inclination will mean direct sight lines to more of the ``inner wall'' of obscuring material closest to the accretion disk.

 Qualitatively it is possible to replicate this postulated relation between L$_{1-5\,\mu m}$/L$_{\rm IR}$ and inclination in the N08 models, with a factor of $\sim2$ increase in the ratio of near-IR to total IR luminosity from inclination 0$^{\circ}-60^{\circ}$ assuming model parameters describing a dense, compact torus ($N\gs10$, $q\gs2$, $Y<10$, $\sigma<30$). Given that the $\sigma$ and $N_{0}$ parameters describe the covering factor (Elitzur et al. 2012) this behaviour in the N08 models is consistent with our geometrically motivated solution; thin (low $\sigma$), dense (high $N$, high $q$, low $Y$) tori should result in large near-IR to total IR luminosity ratios when seen at moderate inclinations ($\gs 40^{\circ}$). Because the opening angle for the torus is fixed at 50$^{\circ}$ in the Stalevski et al.\ (2012) model a similar test for these models cannot be performed.

\section{Conclusions}\label{sec:conc}
We have considered the optical to mid-IR properties of a sample of type 1 quasars selected from a combination of the {\it WISE}, UKIDSS and SDSS datasets.  Using our simple SED modelling approach we estimate a number of quasar properties; the IR-luminosity, the covering factor (from $L_{\rm IR}/L_{\rm bol}$) and the IR SED shape characterised by the ratio of near-IR (1--5\,$\mu$m) to total IR luminosity. Using these measurements we reach the following conclusions:
\begin{enumerate}
\item The distribution of covering factors ($f_C$), defined as the ratio of IR to UV/optical luminosity, is found to obey a log-normal distribution. Once selection limits are taken into account the distribution is characterised by $\langle\log_{10}{f_C}\rangle=-0.41$ and $\sigma_{\log_{10}{f_C}}=0.2$. These values agree well with other IR/radio estimates, but are well below that estimated from comparable X-ray studies.

\item This distribution gives roughly the same shape as that expected for the tilted-disk model (Lawrence \& Elvis 2010), although offset to lower covering factors by $\sim 25$ per cent
\item A significant fraction ($\sim 40$ per cent) of the total IR luminosity is emitted in the near-IR. It is difficult to replicate this behaviour with models in which all the torus material is in ``clumps'' (e.g. Nenkova et al.\ 2008). Smooth or two-phase models, which assume some clumpy material embedded in an otherwise smooth torus, show much higher near-IR to total-IR ratios, consistent with our observations.
  
\item A strong correlation is observed between the ratio of near-IR to total IR luminosity and $f_C$. This is interpreted as a geometric effect, as more of the hotter dust close to the accretion disk will be visible at the high inclinations possible in low $f_C$ quasars.
\end{enumerate}

\section*{Acknowledgements}
We thank the anonymous referee for suggestions which greatly enhanced this work.\\
This publication makes use of data products from the Wide-field Infrared Survey Explorer, which is a joint project of the University of California, Los Angeles, and the Jet Propulsion Laboratory/California Institute of Technology, funded by the National Aeronautics and Space Administration.\\

Funding for the SDSS and SDSS-II has been provided by the Alfred P. Sloan Foundation, the Participating Institutions, the National Science Foundation, the U.S. Department of Energy, the National Aeronautics and Space Administration, the Japanese Monbukagakusho, the Max Planck Society, and the Higher Education Funding Council for England. The SDSS Web Site is http://www.sdss.org/.\\

The SDSS is managed by the Astrophysical Research Consortium for the Participating Institutions. The Participating Institutions are the American Museum of Natural History, Astrophysical Institute Potsdam, University of Basel, University of Cambridge, Case Western Reserve University, University of Chicago, Drexel University, Fermilab, the Institute for Advanced Study, the Japan Participation Group, Johns Hopkins University, the Joint Institute for Nuclear Astrophysics, the Kavli Institute for Particle Astrophysics and Cosmology, the Korean Scientist Group, the Chinese Academy of Sciences (LAMOST), Los Alamos National Laboratory, the Max-Planck-Institute for Astronomy (MPIA), the Max-Planck-Institute for Astrophysics (MPA), New Mexico State University, Ohio State University, University of Pittsburgh, University of Portsmouth, Princeton University, the United States Naval Observatory, and the University of Washington.\\

\label{lastpage}

\end{document}